\documentclass[preprints,article,accept,moreauthors,pdftex]{Definitions/mdpi} 

\firstpage{1} 
\pubvolume{1}
\issuenum{1}
\articlenumber{0}
\pubyear{2021}
\copyrightyear{2020}
\datereceived{} 
\dateaccepted{} 
\datepublished{} 
\hreflink{https://doi.org/} 
\pdfoutput=1

\usepackage[lofdepth,lotdepth]{subfig}
\usepackage{xspace}

\newcommand{\kappad}{\ensuremath{\kappa_d}}
\newcommand{\kr}{\ensuremath{\frac{\kappad}{\kappa}}}
\newcommand{\taud}{\ensuremath{\tau_d}}
\newcommand{\npat}{43\xspace}

\Title{Radiotherapy Effects on Diffuse Low-Grade Gliomas: Confronting
Theory With Clinical Data}

\TitleCitation{Radiotherapy effects for diffuse low-grade gliomas}

\Author{L\'eo Adenis$^{1,2}$, St\'ephane
  Plaszczynski$^{1,2}$*,  Basile 
  Grammaticos$^{1,2}$,  Johan Pallud $^{3, 4, 5}$ and
  Mathilde Badoual$^{1,2}$}

\AuthorNames{Leo Adenis, Stephane
  Plaszczynski,  Basile Grammaticos,  Johan Pallud and  Mathilde Badoual}

\AuthorCitation{Adenis, A.; Plaszczynski, S.; Grammaticos, B.; Pallud ,J; Badoual, M.}

\address{%
$^{1}$ \quad Universit\'e Paris-Saclay, CNRS/IN2P3, IJCLab, 91405 Orsay, France \\
$^{2}$ \quad Universit\'e de Paris, IJCLab, 91405 Orsay France \\
$^{3}$ \quad Department of Neurosurgery, GHU Paris - Sainte-Anne Hospital, Paris, France \\
$^{4}$ \quad Université de Paris, Sorbonne Paris Cité, Paris, France \\
$^{5}$ \quad Inserm, U1266, IMA-Brain, Institut de Psychiatrie et Neurosciences de Paris, Paris, France }

\corres{Correspondence: stephane.plaszczynski@ijclab.in2p3.fr}

\abstract{Diffuse low grade gliomas are slowly growing tumors that always recur
after treatment. 
In this paper, we revisit the
modeling of the tumor radius evolution before and after the
radiotherapy process and propose a novel model that is simple, yet biologically
motivated, and that 
remedies some shortcomings of previously proposed ones.
We confront it with clinical data consisting in
time-series of 
tumor radius for 43 patient records, using a stochastic
optimization technique and obtain very good fits in all the cases. 
Since our model
describes the evolution of the tumor from the very first glioma cell, it gives
access to the possible age of the tumor. Using the
technique of profile-likelihood to extract all the information from
the data, we build confidence intervals for the tumor birth age 
and confirm the fact that low-grade glioma seem to
appear in the late teenage years. Moreover, an approximate analytical
expression of the temporal evolution of the tumor radius allows us to
explain the correlations observed in the data. 
}

\keyword{Mathematical modeling; gliomas; radiotherapy; optimization; data analysis}

\begin{document}

\section{Introduction}
Gliomas are tumors of the central nervous system that arise from
precursors of glial cells and account for almost 80\% of primary
malignant brain tumors. Although relatively rare, they result in more
years of life lost than any other tumor: approximately 13000 deaths
and 18000 new cases of primary malignant brain and central nervous system tumors occur
annually in the US \cite{who07}. 
Historically, the tumors of the central nervous system have been
classified by the World health organization into four grades,
based on their histological characteristics and reflecting the
aggressiveness of the tumor \cite{who07}: grade 1 gliomas are benign, well delineated and can be cured
by surgery. From grade 2 and above, the tumors are diffuse and
because of that, incurable.  
Recently, a revision of the World Health Organization
classification was proposed, reflecting the fact that prognosis is
more strongly associated with molecular diagnostic features than with
histological ones only \cite{who16}. It is now the isocitrate dehydrogenase (IDH)
enzyme mutation status that allows to classify these tumors in the
first place. According to this revision, the group of grade 2
low-grade gliomas now includes diffuse astrocytomas (mutant IDH1) and
oligodendrogliomas (mutant IDH1 and 1p/19q co-deletion). The
highest grade glioma group, that  includes IDH
wildtype glioblastomas and grade 3 astrocytomas, is associated with a life time expectancy of
less than a year \cite{marra19}.  In this paper, we use data from patients who were recruited before 2016 for most
of them. The status of the IDH enzyme mutation was not assessed and
the gliomas were classified according to histological features only.
We have just followed the WHO classification in use at the time of diagnosis and use the term of diffuse low-grade gliomas (DLGG) for these patients' tumors, that
includes low-grade astocytomas and oligodendrogliomas \cite{pallud12a}.

In high-grade gliomas, the rate of proliferation is very large and in the center, the
cells become hypoxic and finally necrotic. In contrast, in DLGG, the
rate of proliferation is lower, and these tumors are composed only of
isolated migrating tumor cells that infiltrate the normal tissue. On a MRI scan, DLGG present a T1 hypointense signal without contrast enhancement (since
there is no angiogenesis) but a T2-FLAIR hyperintense signal \cite{pallud11bis,price10}. It has
been shown that tumor cells migrate well beyond the tumor limits of
this hyperintense area on T2-FLAIR weighted MRI scans  \cite{kelly87, pallud10}.

If DLGG are associated to an extended lifetime expectancy compared to
higher-grade gliomas, they represent a real public health issue since
patients are often young (between 20 and 40 years-old) with a previously normal
social and professional life. DLGG grow slowly but their invasive feature is
responsible for the unavoidable recurrence, even after oncological treatments \cite{pallud10bis}. 

Treatments consist in the first place in surgery when possible. Chemotherapy and then radiotherapy are proposed in front of a progressive residual tumor and at tumor progression. However, despite technical
progress in imaging techniques and therapeutic management,  treatment
only confers a modest improvement in overall survival \cite{ricard07, peyre10, mandonnet09, prabhu10}.
But even worse, all low-grade lesions eventually evolve into higher-grade
malignant tumor, when neoangiogenesis is triggered \cite{bobeck14}.   

For DLGG, the goals of radiotherapy (RT) are to control tumor growth, improve
progression-free survival and patient quality of life by reducing the
risk of seizures, and delay anaplastic transformation \cite{vandenbent05}. 


Several aspects of DLGG have already been the object of models, from their origin \cite{dufour18} to their natural evolution \cite{mandonnet09, gerin12}, their response to treatments (in particular with RT \cite{ribba12, badoual14, perez15,  galochkina15, boudia19}) and their anaplastic transformation \cite{bogdanska17}. The diffusion-proliferation model plays a
special role in the modeling of glioma evolution field. It is based on
a differential equation governing glioma cell density, and in its
simplest form, involves only two key phenomena (and thus two
parameters): the migration (modeled as a diffusion) of the cells and
their proliferation. This model has been extensively used for high-grade gliomas  \cite{harpold07, badoual10, corwin13, unkelbach14,amelot17}, but in fact, the model is more adapted to DLGG. Despite its simplicity, this model can
reproduce in particular an important feature of DLGG growth that has been verified with clinical data, which is that the tumor radius increases linearly with time (at large time) \cite{mandonnet03, pallud06}. 

The most striking feature of
the evolution of DLGG under RT, that any model should reproduce, is
that the tumor radius continues to decrease even after the end of
the treatment. This delay can range
form a few months to almost ten years, depending on the patient, before the tumor systematically recurs, and starts to grow again. In
\cite{ribba12} two populations of cells are defined, one that is
quiescent and the other one that proliferates. RT damages the cells of
the proliferating population, thus transforming them into quiescent
cells. The model is based on ordinary differential equations and does
not include any spatial structure. However, in the case of gliomas,
the spatial structure is essential in a model since a key feature of
DLGG is their capacity to invade surrounding normal tissue. In
\cite{rockne10} the authors use the diffusion-proliferation model, with cell
death term due to RT (present only while the therapy lasts), and apply
it to high-grade gliomas. But this model is not adequate for low-grade
gliomas since it cannot account for the most striking feature of the
clinical follow up, i.e. the reduction of tumour radius that lasts much
longer than the treatment by RT itself. In \cite{badoual14} our group proposed a diffusion-proliferation model coupled with
the production of edema by tumor cells. We fitted successfully 29
follow-ups of patients. However, even if this model is the closest to
the biological characteristics of DLGG, it involves 5 parameters,
including the two parameters for the edema dynamics. These two
parameters are unknown and cannot be easily measured experimentally.
Without any estimation of their values to compare with, it is
difficult to validate the model and make predictions. 
In \cite{perez15}, the authors developed a model based on the
diffusion-proliferation model that involves two cell populations, the damaged by RT and the not damaged one, close to the one in \cite{ribba12}. The
advantage of this model is that it contains a spatial structure and
also allows a slow decrease of the tumor radius after the end of the
RT treatment. The authors used the model to study the impact of a
fractionation of the RT treatment \cite{galochkina15,henares17}. However, these studies are
theoretical and the model was not applied to real clinical data. 

In this article, we develop a simple biophysical model of DLGG evolution based
on the diffusion-proliferation model with the addition of the effect
of RT and confront it to a large number (43) of patients clinical data.
We use state-of-the art analysis techniques to adjust the model 
and show that it is possible to get an excellent agreement between the model and the data, for all the patients. We then study the birth age
of the tumors, the parameter values and the correlations among several
observables before and after RT.

\section{Materials and Methods}

\subsection{The patients} 
\label{sec:patients}

We had at our disposal a set of \npat patients with DLGG, diagnosed at
the Sainte-Anne Hospital (Paris, France) from 1989 to 2000. These
patients were selected according to precise criteria detailed
elsewhere \cite{pallud12a}. In short, only adults with typical DLGG
(that is, no angiogenesis and thus no contrast enhancement on
gadolinium- T1 images), available clinical and imaging follow-up
before, during, and after RT and RT as their first oncological
treatment except for stereotactic biopsies, were eligible. 
The patients had MRI follow-up before, during and after RT. Three
tumour diameters in axial, coronal and sagittal planes on each MRI
image with T2 weighted and FLAIR sequences were measured manually. The
mean radiological tumour radius is defined as half the geometric mean
of these three diameters and is measured as a function of time. The
error bars for the measured mean radius have been estimated by
clinicians and are set to $\pm 1$ mm. From this cohort, we discarded the patients that didn't have any sign of tumor regrowth at the last time point, or those that have fewer than five time points in their follow-up.

\subsection{The model}
\label{sec:model}

\subsubsection{Free tumor evolution}
\label{sec:free}
The diffusion-proliferation model describes the evolution of the glioma cell density $\rho$ as 
\begin{equation}
\label{eq:freemodel}
{\partial \rho \over\partial t}=D\Delta\rho+\kappa\rho(1-\rho)
\end{equation}

where $\rho(\vec{r},t)=C/C_m$, $C$ being the glioma cell density and
$C_m$ the maximal cell concentration that the tissue can handle (also
called the carrying capacity), $D$ is the diffusion coefficient of the
glioma cells, and  $\kappa$ is the proliferation coefficient.  

A tumor is a 3D object so it seems logical to solve Eq (\ref{eq:freemodel}) in
3D. We do not want to enter into too many details about its precise shape
for each patient so we will assume a spherical symmetry of all tumors.

In 3D, assuming a spherical symmetry of the tumor, Eq (\ref{eq:freemodel}) becomes:

\begin{equation}
\label{eq:evolve1D}
{\partial \rho(r,t) \over\partial t}=D({{\partial}^2 \rho(r,t)\over{\partial r}^2}+{2 \over r} {\partial \rho(r,t) \over{\partial r}})+\kappa\rho(r,t)(1-\rho(r,t))  
\end{equation}

As explained in \cite{gerin12}, when introducing an auxiliary variable
$u=r \rho$, Eq (\ref{eq:evolve1D}) takes the form: 

\begin{equation}
  \label{eq:evolve3D}
{\partial u \over\partial t}=D{{\partial}^2 u \over{\partial r}^2}+\kappa u(1-{u\over r})  
\end{equation}
with $u(r=0,t)=0$ and ${\partial \rho \over \partial r}(r=0,t)=0$.

We solve Eq (\ref{eq:evolve3D}) by discretizing it on a mesh of spatial size
$\delta r=10^{-2}$ mm=$10~\mu$m and with a time step $\delta t=10^{-2}$ yr,
using an implicit scheme for the diffusion part and a homographic-type
discretisation for the logistic part.  

The limit of MRI-signal abnormality (with T2-weighted or FLAIR
sequences) is usually assumed to be a curve of iso-density of glioma
cells. The radius of this visible part of the tumor on MRI (usually
called the ``tumor radius") is thus defined as the distance $r$ to the
tumour centre where the cell density $\rho$ crosses a fixed threshold
$\rho^*$. The value of this parameter $\rho^*$ is not known precisely,
but we expect that its value, as long as it stays much smaller than 1,
will not have a strong influence on our conclusions. We will set
$\rho^*=0.02$ for all the simulations \cite{harpold07, perez15}.  

The initial conditions are the same for all the simulations and
correspond to the appearance of the first tumor cell:  $\rho(r = \delta r,0)=1$
and $\rho(r>\delta r,0)=0$. Here, we assume that the tumor has been
developing with the same proliferation and diffusion coefficient
since the appearance of the first tumor cell.

\subsubsection{Modeling RT} 
\label{sec:RT}
Next we turn to the modelling of the radiotherapy process itself. The
action of RT on the glioma cells was modeled as an instantaneous event
since the duration of the treatment (typically 6 weeks, or 0.11~yr) is
negligible compared to the mean regrowth delay after RT (1.25~yr for
our patients) \cite{pallud12a}. The origin of time is set to the time of
RT.  

We introduce a new model to capture the essence of what happens
after the radiotherapy, by adding 
to the free evolution Eq.(\ref{eq:freemodel}) a
\textit{time-dependent} death term: 

\begin{equation}
  \label{eq:model}
  {\partial \rho_(\vec{r} ,t)\over\partial t}=D\Delta\rho(\vec{r},t)+[\kappa-\kappa_D(t)] \rho(\vec{r},t) (1-\rho(\vec{r},t)).
\end{equation}
The simplest way to introduce some characteristic time is to choose
\begin{equation}
\label{eq:kappad}
 \kappa_D(t)=\kappad e^{-(t-t_r)/\tau} 
 \end{equation}
for $t>t_r$ where $t_r$ is the time of RT
 and  $\kappa_D(t)=0$  for $t<t_r$.

To the two parameters that describe the natural evolution of the
tumor ($\kappa, D$), and the two others related to the effect of RT on
tumor cells ($\kappad, \tau$), we add a fifth one, the tumor age $T$
at the time of RT. Although not derived from physical modeling it is an
unknown of the problem that must then be determined with the others
\footnote{In statistical terminology it is a \textit{nuisance} parameter.}. This parameter is important because we
need to ensure that $T$ is always smaller that the age of the patient
himself at the time of RT.

\subsection{Fitting procedure}
For each patient we determine the set of parameters that best fits our
data by performing numerically a multidimensional minimization of the
objective function:
\begin{equation}
  \label{eq:chi2}
  \chi^2(T,D,\kappa,\kappad,\taud)={\displaystyle \sum_{i=1}^{N_{data}}} \left[R_{data}(t_i)-R_{mod}(t_i; T,D,\kappa,\kappad,\taud)\right]^2,
\end{equation}
 
where $R_{data}(t_i)$ denotes the measured radii at time $t_i$
and $R_{mod}$ is the theoretical model that is numerically computed by
evolving the cell concentration profile following our model equations 
and thresholding it at $\rho^\ast$ to obtain the radius \footnote{We
  recall that the error on the measurements is about 1 mm so there is
  no need to rescale the residuals.}.

We also add the constraint that from a radius of 15 mm, the 
the tumor should evolve almost in the asymptotic regime. This linearity has been observed in clinical data \cite{mandonnet03} and has already been implemented in  \cite{gerin12}.
More specifically we compute the relative difference of the speed of the
model with respect to the asymptotic value $c=2\sqrt{D \kappa}$
\begin{equation}
  r_{15}=\left(\frac{d R_{mod}}{dt}~ (15)-c\right)/c,
\end{equation}
and if this value exceeds 20\% we add to the $\chi^2$ a quadratic term
\begin{equation}
  \chi^2_{extra}=\left(\frac{r_{15}-0.20}{0.01}\right)^2.
\end{equation}

Finally, to avoid aberrant values we will use some light bounds on the
possible parameter range $0<D<10 ~\rm{mm}^2/\rm{yr}$, $0<\kappa<10
~\rm{yr}^{-1}$ , $0<\kappad<500 ~\rm{yr}^{-1}$ and $0<\taud<50~ \rm{yr}$.

The 5D optimization problem from a non analytical and non linear equation is
challenging for standard minimization procedures that often rely on the use of analytical gradients, but which are not available here.
After several tests, the optimization method we have chosen is the
\textit{covariance matrix
adaptation evolution strategy} (CMA-ES\footnote{\url{http://cma.gforge.inria.fr}}) \cite{CMA} . It is a stochastic method that
belongs to the class of evolutionary algorithms, and is often used
for challenging optimization problems.
The algorithm CMA-ES proceeds as follows: at each time step, several
new candidate solutions are sampled from a multivariate normal
distribution and the $N$ candidate solutions that correspond to the
smallest value of the objective function $f$ are selected. A weighted
combination of the $N$ best candidate solutions is used to update the
internal state variables such as the mean of the distribution of
candidates, the step-size and the covariance matrix.  
One advantage of this method over other evolutionary ones is that
there are only a few parameters that have to be chosen: the starting
point, some estimate of the associated errors (which we choose to be
about 10\%) and the population size that we tuned to 50 to obtain
stable results. For each patient, since the algorithm is stochastic, 10 runs are
performed and the best fit (lowest $\chi_{min}^2$ value) is kept. In
practice the 10 results are very similar.

\section{Results}
\label{sec:results}

\subsection{Characterisation of our model} 
\label{sec:carac}
\begin{figure}[H]
  \centering
  \includegraphics[width=10.5 cm]{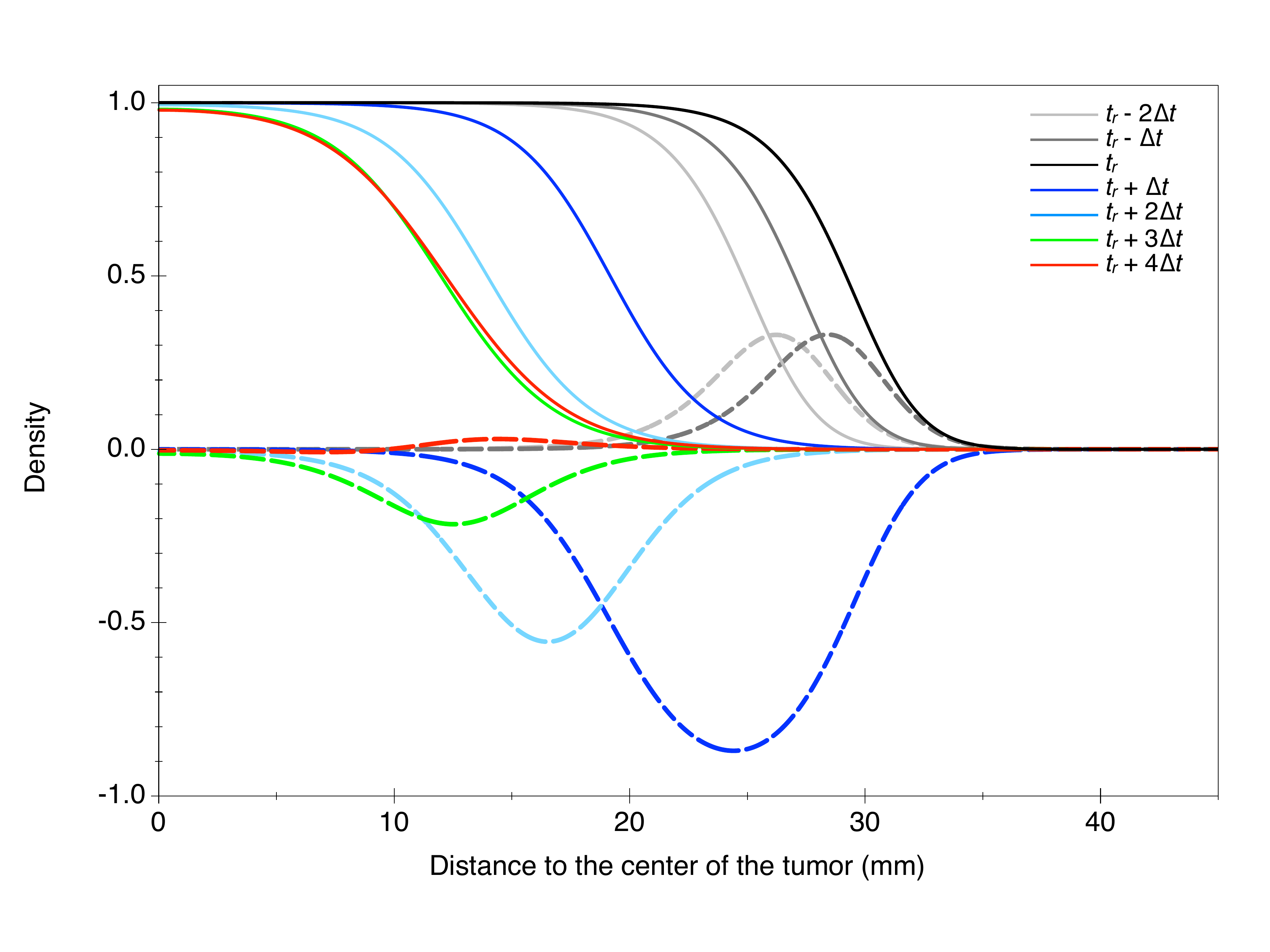}
  \caption{Cell density (full lines) and proliferating/killed cell
    density (dashed lines) profiles for different times, before (light and dark gray lines), at the time of RT (black lines) and after RT (coloured
    lines). The time interval between two profiles is $\Delta t=1.1$~yr. The
    profile of proliferating/killed cells is obtained by substracting
    two successive profiles. Parameters: $\kappa = 1.3$~yr$^{-1}$, $D = 0.8$~mm$^2$~yr$^{-1}$, $\kappa_d=8$~yr$^{-1}$,$\taud=2$~yr.}
  \label{fig:figprolif}
\end{figure}

In Eq (\ref{eq:model}), the $\kappa-\kappa_D(t)$ term accounts for a net proliferation
that can be positive if cells are actually created (before RT, dashed
lines on Figure \ref{fig:figprolif}), or negative if cells are killed
(after RT, coloured lines).  On Figure \ref{fig:figprolif}, one can see that before RT, the
front of the profiles moves with a constant positive velocity and the
same amount of proliferating cells is created during a given time
interval (light gray, dark gray and black profiles of proliferatiing cells, dashed lines). Since the center of the tumor reaches saturation, the
proliferating cells are located at the border of the tumor. After RT,
the front moves backwards, and the net proliferation becomes negative:
cells are killed. Since the death term has exactly the same structure
as the proliferation term, cells do not die where the cell density is
close to saturation, at the center of the tumor. Cells are
killed at the border, and because the death parameter decreases
exponentially with time, the amount of cells dying during each time
interval decreases. After some time, proliferation surpasses death and
the tumor starts to regrow (see the pink profile,
Figure \ref{fig:figprolif}).

This is different from models with two populations (dammaged/undamaged
cells) with constant death rates where the density is uniformly
decreased. As shown later these models lead to a linear
decrease just after RT (which is clearly not what is observed), while our
model allows for an exponential-type decrease (more about the comparison
in the Appendix).

Simple analytical considerations can give insights about the
linear-versus-exponential early decrease of the radius.  
If we assume that at the time of RT the asymptotic regime is reached,
then the profile of the cell density is a sigmoidal curve and the
front propagates at a constant velocity: 
\begin{equation}
  \label{eq:v}
v=2\sqrt{D\kappa}   
\end{equation}

At the time of RT ($t=t_r$), the profile of the cell density then
follows \cite{murray-livre-tome-2}: 
\begin{equation}
  \label{eq:murray}
\rho(r,t_r)\simeq {1 \over 1+\exp({(r-r_{1/2})/\lambda})}  
\end{equation}
where the characteristic length is $\lambda=2\sqrt{D \over \kappa}$
and $r_{1/2}$ is defined so that $\rho(r_{1/2},t_r)=1/2$.

Since we are interested in the evolution of the radius, corresponding to a
very low threshold of cell density ($\rho^*=0.02 <<1$) the profile is
locally well described near $\rho^\ast$ by
\begin{equation}
\rho(r,t_r)\simeq {\exp({-(r-r_{1/2})/\lambda})}.
\end{equation}

Just after RT, the time during which the radius decreases before
regrowth is short enough for us to neglect the effect of the diffusion on the shape of the
front.  
In this case the cell density follows the equation: 
\begin{eqnarray}
  \label{eq:approx}
  {d\rho\over dt}& =(\kappa-\kappa_D(t)) \rho (1-\rho) \nonumber \\
  & \simeq (\kappa-\kappa_D(t)) \rho   
\end{eqnarray}
near the threshold where the saturation term can be neglected.

The solution to this equation for $t>t_r$ is:
\begin{equation}
\rho(r,t)=\rho(r,t_r)\exp{(\kappa (t-t_r)-\int_{t_r}^t \kappa_d dt)}.  
\end{equation}

So after RT the cell density close to the threshold (at large $r$),
can be rewritten as:
\begin{equation}
\rho(r,t)=\exp{(-(r-r_{1/2})/\lambda+\kappa (t-t_r)-\int_{t_r}^t \kappa_d dt)}.  
\end{equation}

Setting $\rho=\rho^*$, one finds the evolution of the radius at the threshold:
\begin{equation}
r(t)=r_{1/2}+\lambda(\kappa (t-t_r)-\int_{t_r}^t \kappa_d dt-\ln(\rho^*)).   
\end{equation}

If the death term is constant, then $\int_{t_r}^t
\kappa_d dt=\kappa_d (t-t_r)$ and the radius just after RT varies \textit{linearly}
with a constant velocity: $\lambda(\kappa-\kappad)$.
On the other hand, if $\kappa_D(t)$ is an exponential function, we obtain: 
\begin{equation}
  r(t)=r_{1/2}+\lambda(\kappa (t-t_r)+\kappad\taud (e^{-(t-t_r)/\taud}-1)-\ln(\rho^*)).
\end{equation}

We can write this equation in a more simple way, by reintroducing the effective velocity $v$ (Eq \ref{eq:v}) and defining $r_r=r(t_r)$ being the radius of the tumor at the time of RT. After RT (for $t>t_r$) the equation of evolution of the tumor radius is:
\begin{equation}
  \label{eq:rast}
  r(t)=r_r - v \taud {\kappad \over \kappa}(1-e^{-(t-t_r)/\taud})+v(t-t_r)
\end{equation}

This evolution is similar to a two-population model where a damaged one
decreases exponentially at some characteristic time $\taud$ and amplitude $v \taud {\kappa_{dm} \over \kappa}$ and
an undamaged one that still grows linearly at the asymptotic speed
$vt$. 

Finally we stress that this is only an approximate description aiming at 
capturing the gross features of the radius evolution but 
that in the following the exact Eq (\ref{eq:model}) is numerically solved.

\subsection{Best fits}
\label{sec:bf}
\begin{figure}[H]	
\widefigure
\includegraphics[width=15 cm]{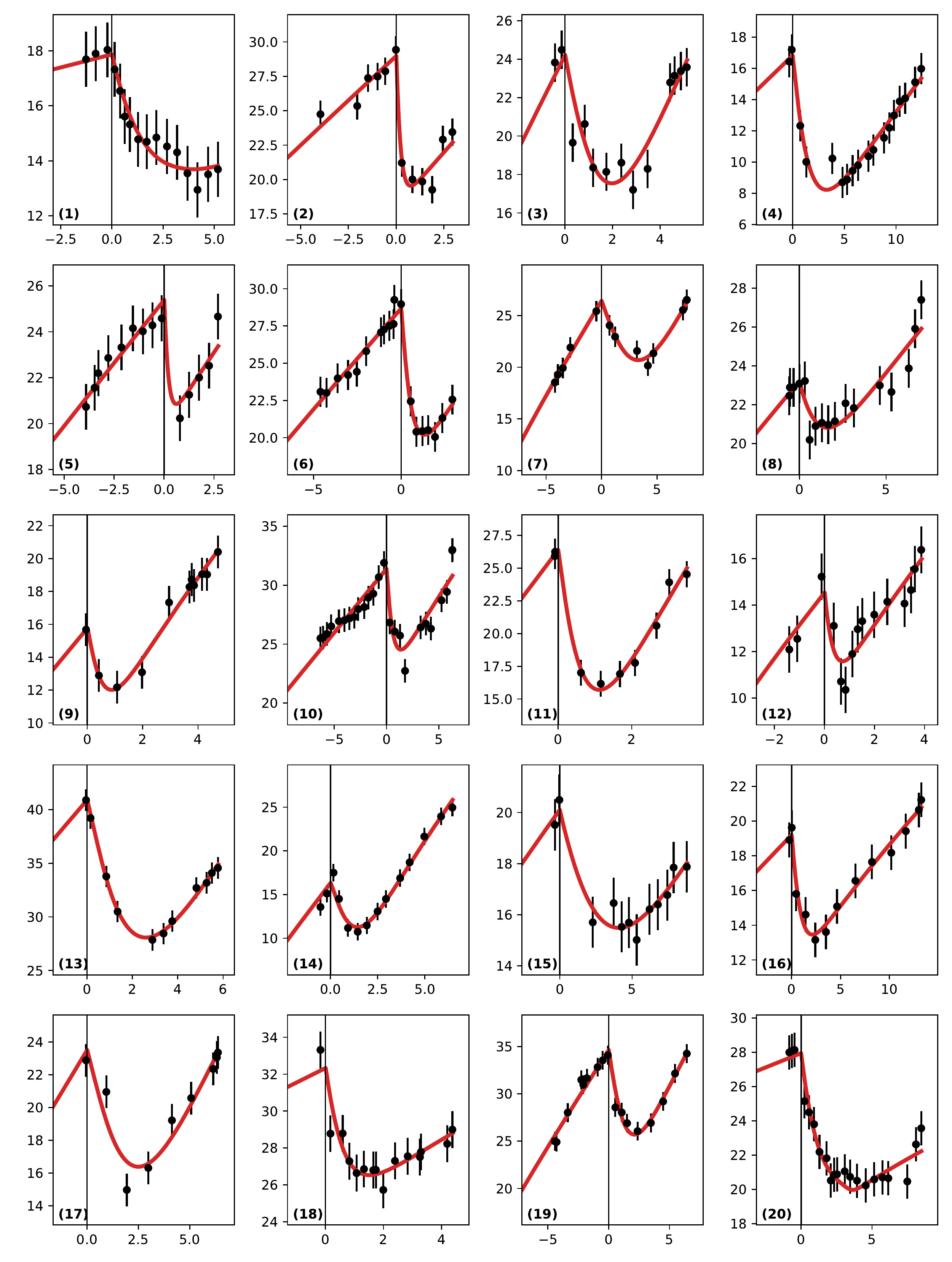}
\caption{Comparison between the data points (in black) and our best
    fit models (red curve) for 
    20 patients. The abscissa represent time in years (with the origin at RT)
    and the ordinate the tumor radius (in mm). Note the scales are
    floating and span various ranges. The errors bars on the measurements
    are of 1 mm.\label{fig:bf}}
\end{figure}

For each of our \npat patients, we performed the 5D minimization of the
Eq (\ref{eq:chi2}) objective function; each one leads to a minimal value of
the $\chi^2$ function (called $\chi^2_{min}$ ) for a set of parameters
$(\hat T,\hat D,\hat \kappa,\hat \kappa_{d},\hat \tau)$ that
represents the "best fit".
Figure \ref{fig:bf} shows the agreement between our best fit model and
the data for a large number of patients with various medical follow-ups.

\end{paracol}

\newpage
\begin{paracol}{2}
\switchcolumn


Our model allows to reproduce all the various cases in a very satisfactory way and
although for space reasons we present results for 20 patients (the ones
with the largest number of points) this is the case on all our \npat
datasets (the 23 remaining fits are available in the
  Supplementary Material).




\subsection{Tumor age}
An original aspect of this work is that we consider the age of the
tumor (defined with respect to RT time) as a free parameter, and we will now show that  it is possible to get some information about this parameter.

Even when there is a substantial number of points before RT (which
is rarely the case), one cannot simply extrapolate linearly back in
time to determine the tumor birth date since
there exists an invisible phase corresponding to the development of the tumor, but below the detection level and thus not detectable
\cite{gerin12, pallud13b}. In order to put some constraint on the tumor age $T$, we resort to the
technique of the \textit{profile-likelihood} (see e.g.. \cite{james})
that works in the following way.

The tumor age is \textit{fixed} at some value $T$ and a
minimization over the 4 remaining parameters is performed, giving a
$\chi^2_{min}(T)$ value. The procedure is repeated for several $T$
values. Taking as zero point the lowest value, one can reconstruct the profile-likelihood $\Delta \chi^2(T)$ of the tumor age,
which can now be used to put a quantitative constraint on the $T$ parameter.
Indeed it can be shown (e.g. \cite{ward}) that this function converges to a $\chi^2$
distribution with one degree of freedom so one can use its
quantiles to get confidence level intervals. In particular one obtains
a 95\% confidence level intervals by thresholding the profile
likelihood at 3.84.

We have reconstructed the constraint on the tumor age (at RT) for all our selected 20 patients
\footnote{they are available as Supplementary Material} and highlight some
typical cases that show why the constraint depends crucially on the
data on the patient follow-ups (Figure \ref{fig:bf}).

\begin{itemize}
\item{(a):} patient (6): this is a case  where there
  are many points before RT and few during the regrowth phase.
\item{(b):} patient (14) is the inverse, few points before RT but the
  regrowth is strongly sampled.
\item{(c):} Patient (13): no points before RT and a few during regrowth.
\end{itemize}

The corresponding profile-likelihoods are shown on Figure \ref{fig:profall}.

\begin{figure}[H]
  \centering
  \subfloat[]{\label{fig:prof_a} \includegraphics[width=4.5 cm]{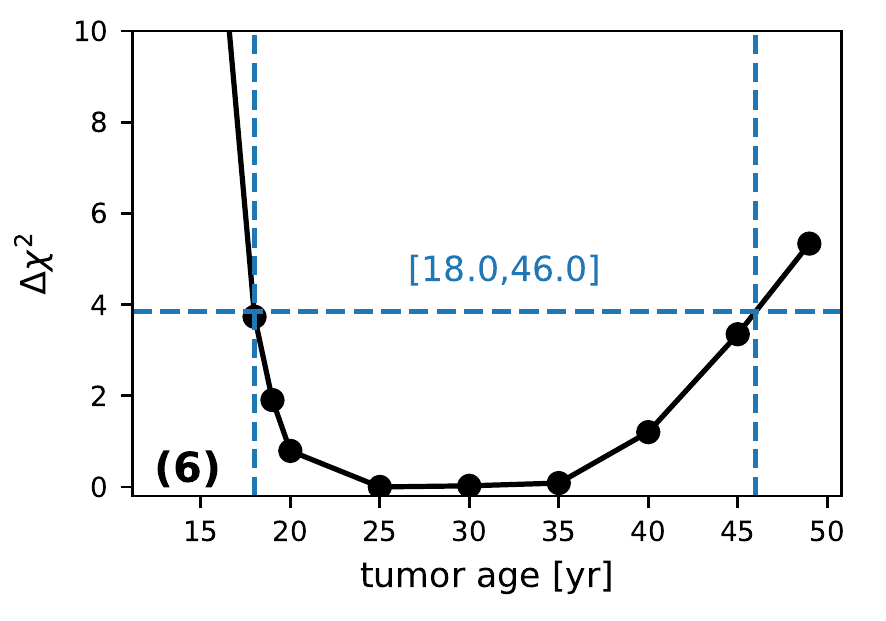}}
  \subfloat[]{ \label{fig:prof_b} \includegraphics[width=4.5 cm]{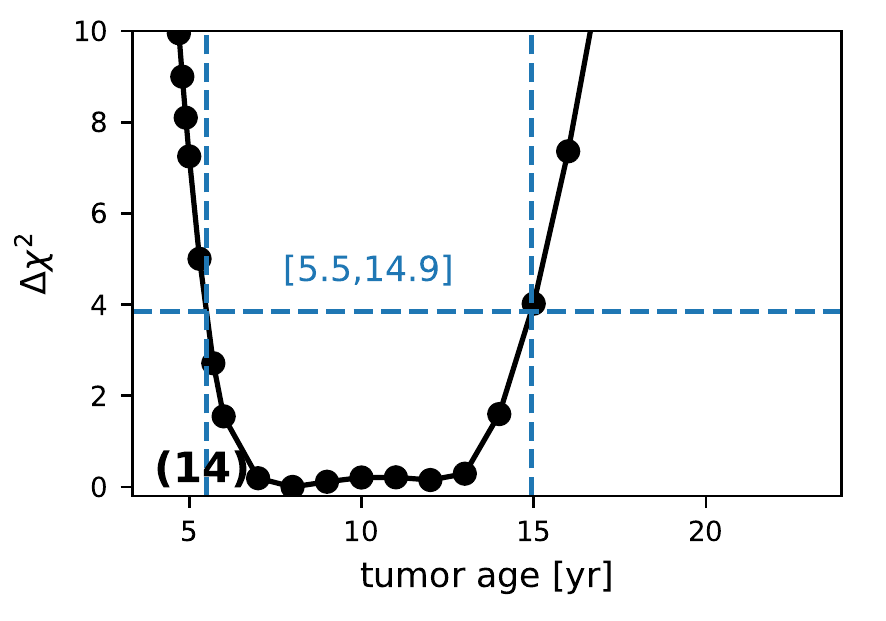}}
  \subfloat[]{\label{fig:prof_c}  \includegraphics[width=4.5 cm]{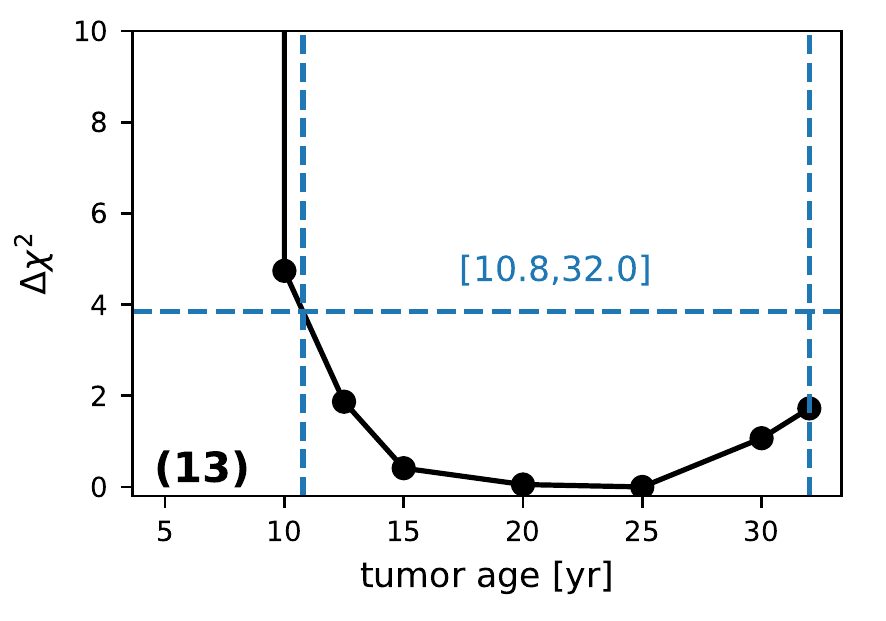}}
  \caption{Constraints on the tumour age (at RT) with the 
     profile-likelihood method described in the text for 3 patients. 
     Considering the region below
     $\Delta \chi^2<3.84$ (the horizontal dashed line) one reconstruct a 95\%
     confidence level interval (shown by the dashed
     vertical lines and corresponding range values). The last value of
   the abscissa is the patient age.}
  \label{fig:profall}
\end{figure}

For patient (a) one obtains both a minimum age constraint (18 years)
and a maximum one (46 years). The points before RT fix both
the radius at RT and the slope. The latter essentially constrains the  product $D \kappa$ through the asymptotic speed ($v=2\sqrt{D\kappa}$).
The invisible phase depends essentially on the proliferation rate
$\kappa$ and has a natural limit \cite{gerin12}. Its duration cannot
be smaller that the time at which the first point was measured. This fixes the tumor age lower limit.
The upper limit comes from the fact that when the age increases, the
proliferation rate is getting smaller. For small $\kappa$ the radius
evolution after the invisible phase is more and more curved which
finally is in disagreement with the data near RT and our constraint on the linearity of the evolution at $r=15$ mm. This fixes the tumor age
upper limit.
Between those limits, several models, corresponding to several sets of parameters, fit equally
well the data. This is illustrated on Figure \ref{fig:faisceau} where we
plotted the four models corresponding to the bottom points of
Figure \ref{fig:prof_a} (at 20, 25, 30 and 35 years) which all agree well
with the data. One can see that the black model, the farthest from RT (35
years) has more curvature than the others and that the red one, with
the smallest age (20) corresponds to a very brief invisible phase. In this case, the silent phase is close to its minimum compatible wit the data points for this patient.

\begin{figure}[htbp]
   \centering
     \includegraphics[width=10.5 cm]{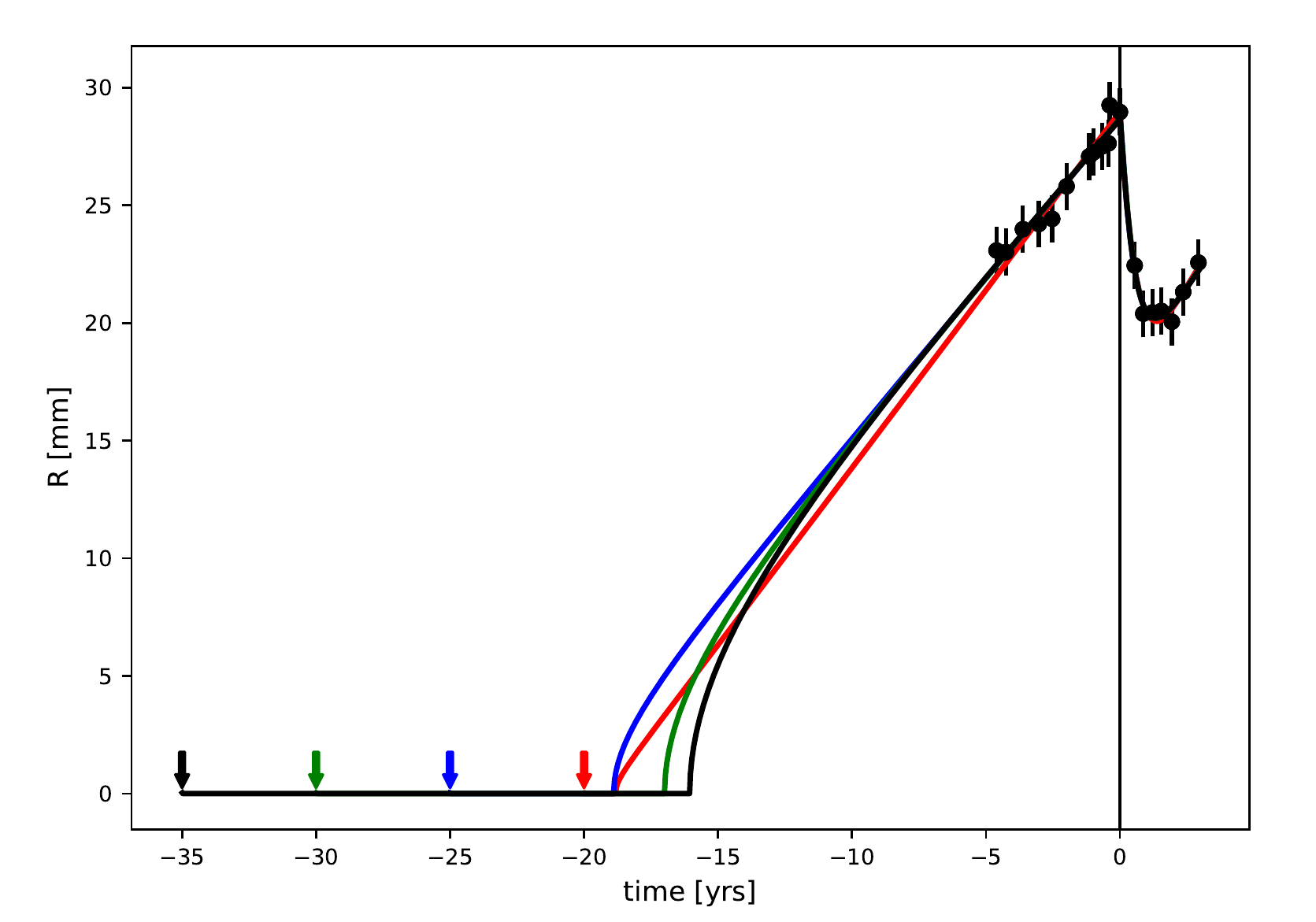}
   \caption{Full radius evolution of the 4 models corresponding to the
     bottom points ($\Delta\chi^2<1$) of Figure \ref{fig:prof_a}. The time origin is fixed at RT and 
     the starting time of the tumor is indicated by colored arrows.
     All the models give 
     similar $\chi^2$ values with respect to the data.
    }
   \label{fig:faisceau}
\end{figure}

Patient (14) does not have much constraint before RT
(Figure \ref{fig:bf}) however the same type of constraint arises from the
regrowth phase that still follows the asymptotic limit, so we still
obtain a full range of valid tumor ages.

Finally, although patient (13) has no points before RT and has points only at
the beginning of the regrowth phase, one can still put a lower limit
on the tumor age through the full fitting of the five parameters model to the data.
This demonstrates the potential of this method that can still put some
minimum bound on the tumor age by exploiting the full data information.
 
From the profile-likelihood reconstruction of all the 20 patients we
can study the age of the patients at the \textit{birth} of the tumors, which is calculated as the age of the patient at the RT minus the age of the tumor at RT.  
We show on Figure \ref{fig:birth} all the 95\% CL intervals obtained with
this method. Although the constraints depend crucially on the data (size and
sampling dates) they are consistent with a DLGG appearance at adolescence, as predicted 
in \cite{gerin12}. 

\begin{figure}[H]
  \centering
       \includegraphics[width=10.5 cm]{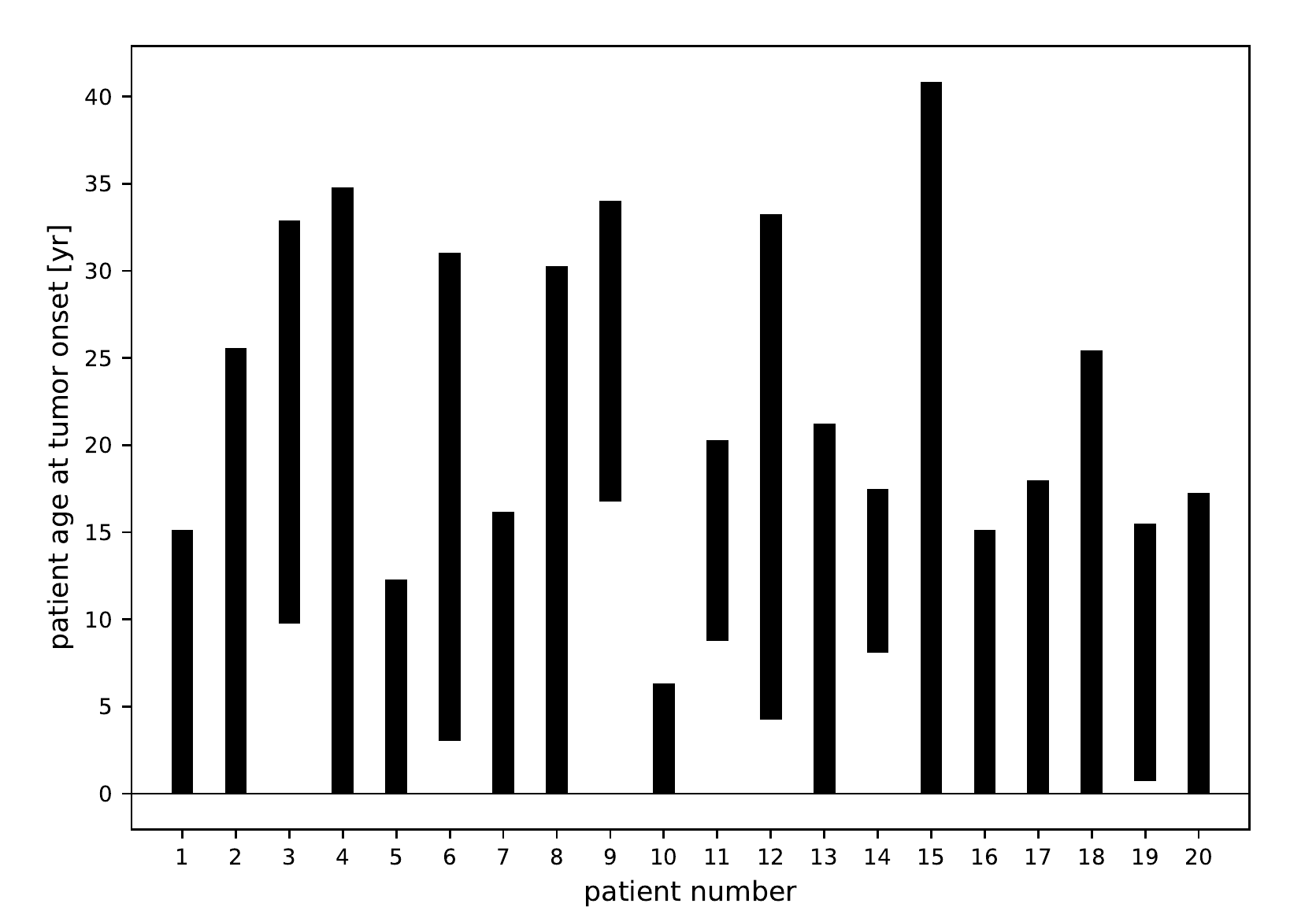}
  \caption{95\% confidence level intervals of the age of the patients at the onset of the tumor, determined from the profile-likelihood analysis.}
  \label{fig:birth}
\end{figure}

\subsection{Tumor characteristics}
\label{sec:parameters}

As already pointed out, the tumor age is a parameter that is very
different from the other ones: it is only an unknown of the
problem, this is why we have treated it separately.
The other four parameters ($D,\kappa,\kappad,\taud$) 
describe the DLGG evolution but, for a given tumor age, they are
strongly correlated.

To illustrate it, let us consider again all the models corresponding to
the ``bottom points" of Figure \ref{fig:prof_a} ($\Delta
\chi^2<1$). We show their values depending on the tumor age on Figure \ref{fig:scan_a}.
While all the models are essentially equivalent in terms of $\chi^2$
the parameters vary considerably (in a correlated way) prohibiting any
interesting individual constraint.

Inspired by Eq (\ref{eq:rast}), we propose to use instead the following
parameters: $v=2\sqrt{D\kappa},\kr$ and $\taud$. We show on Figure \ref{fig:scan_b}
that they are indeed stable in the valid age range so that the
variables are now uncorrelated.

\begin{figure}[H]
  \centering
  \subfloat[]{
    \label{fig:scan_a}
    \includegraphics[width=7cm]{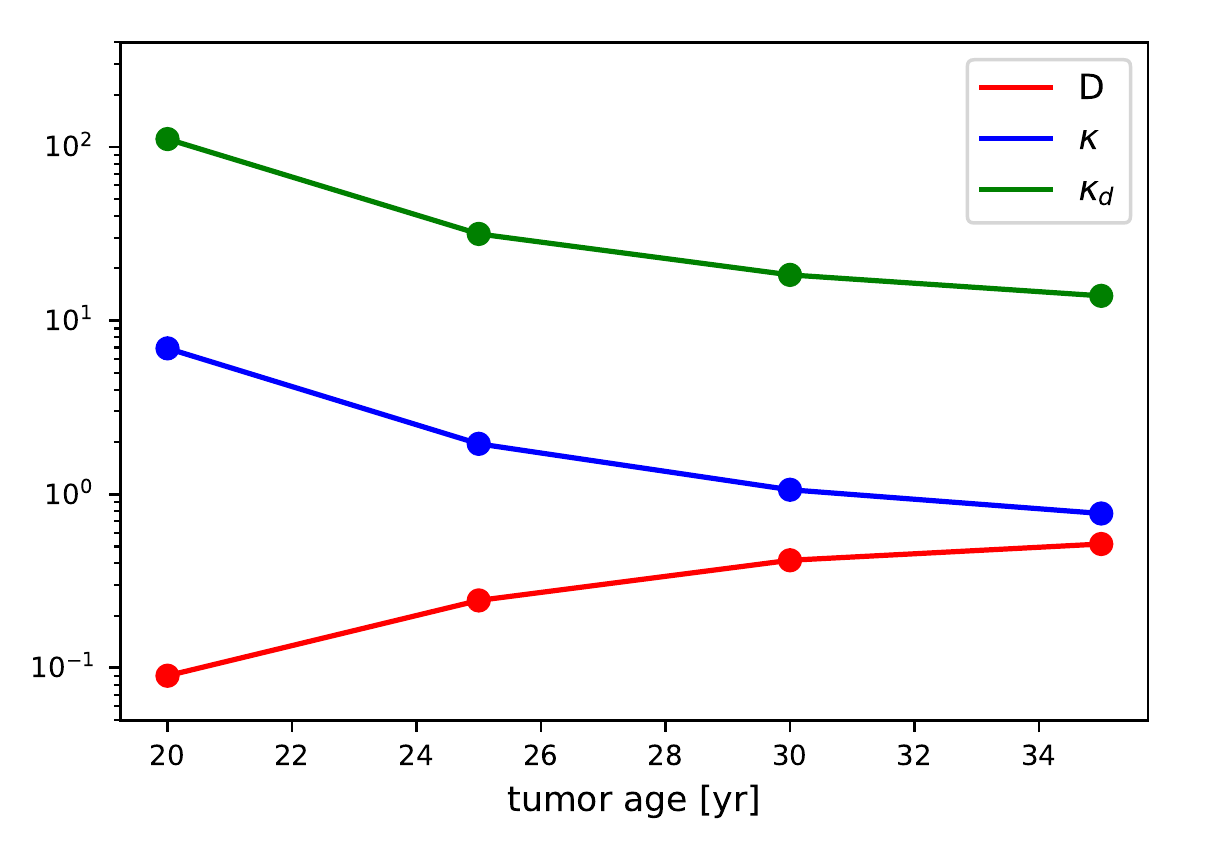}
    }
  \subfloat[]{
    \label{fig:scan_b}
    \includegraphics[width=7cm]{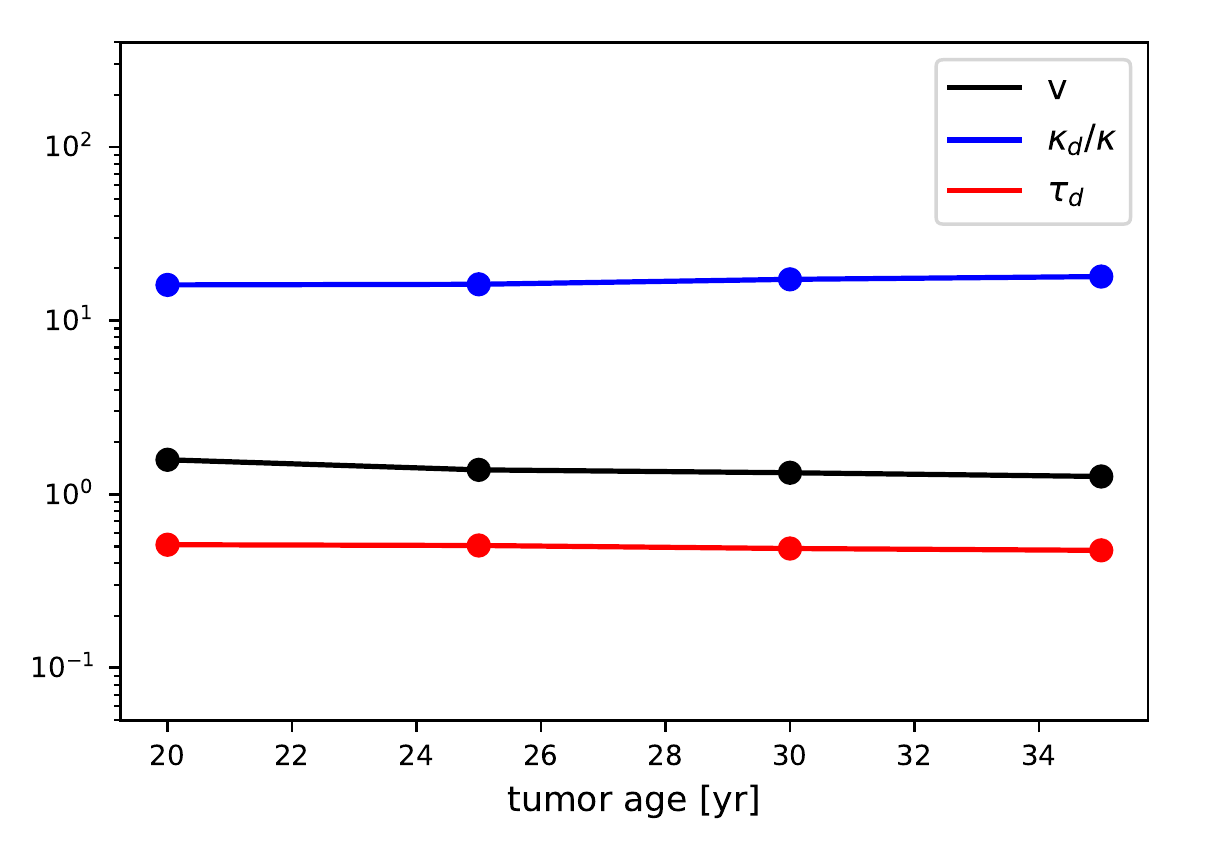}
  }
  \caption{(a) Value of diffusion, proliferation and death-rate
    coefficients of the 4
    models corresponding to Figure\ref{fig:prof_a}
    bottom points ($\Delta\chi^2<1$) . (b) same for the transformed
    set of variables described in the text.}
  \label{fig:scans}
\end{figure}

We then study if there are some common features among our patients.
Figure \ref{fig:histos} shows the histograms of the measured characteristics.

\begin{figure}[H]
  \centering
  \subfloat[]{
    \includegraphics[width=4.5cm]{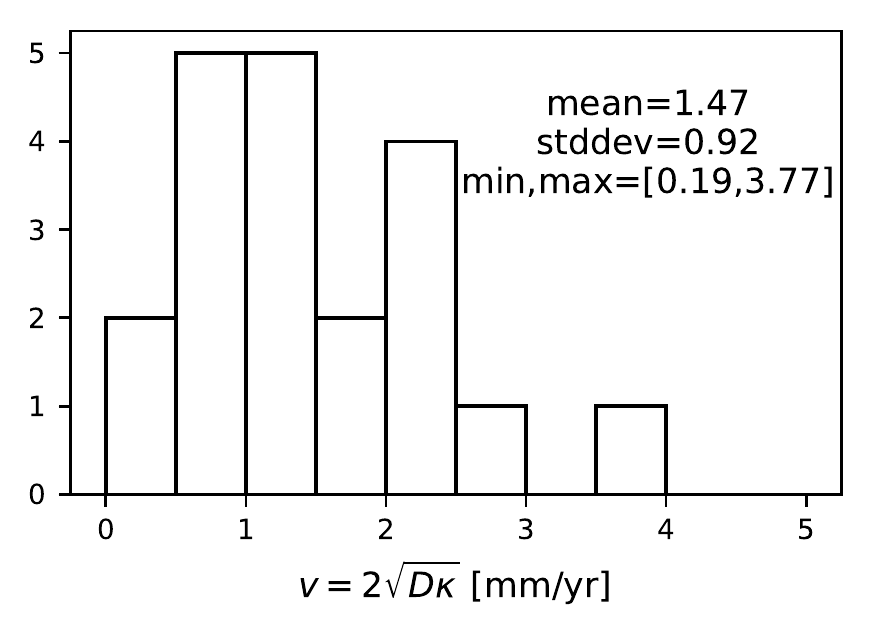}
    }
  \subfloat[]{
    \includegraphics[width=4.5cm]{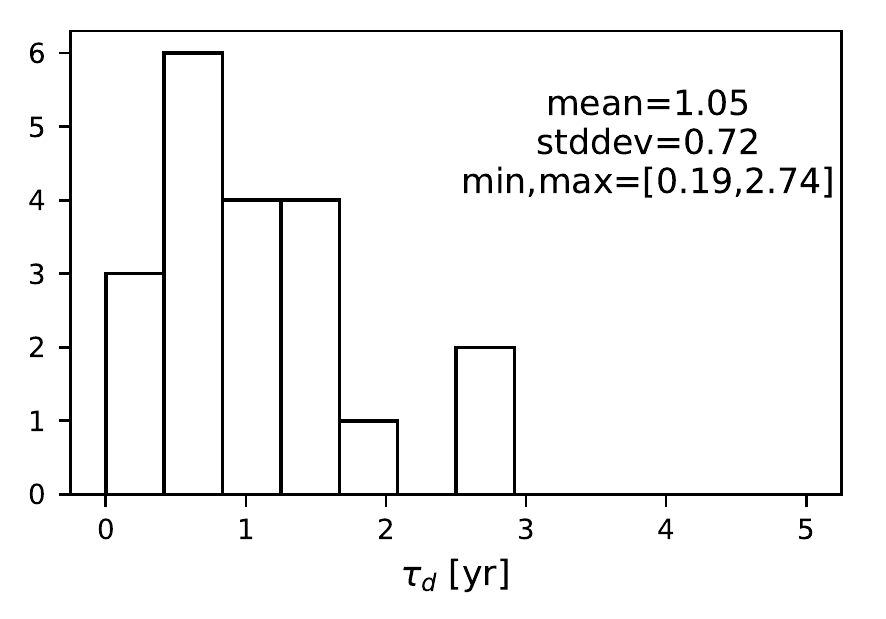}
  }
  \subfloat[]{
    \includegraphics[width=4.5cm]{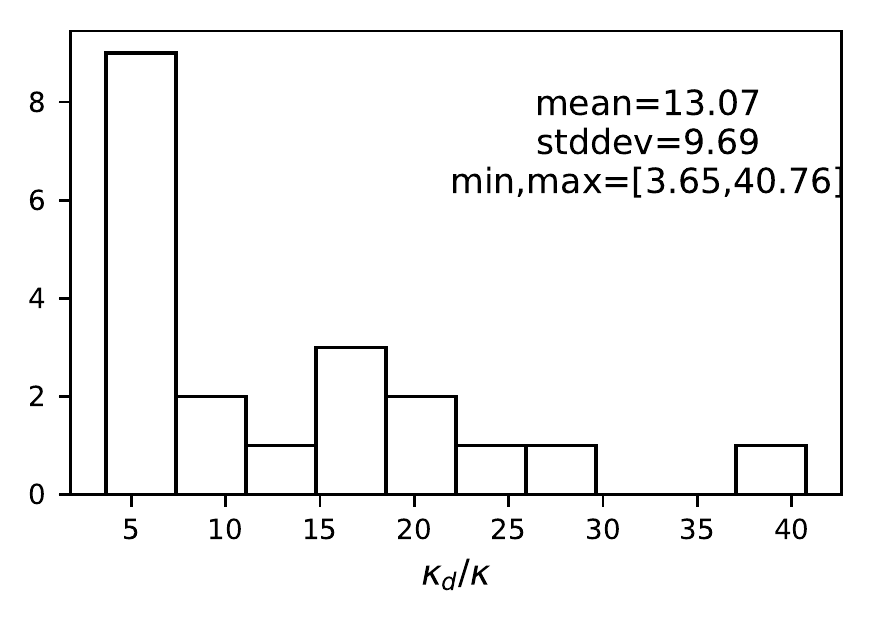}
  }
  \caption{Histograms of the best-fit parameters
    for the velocity
    parameter (a), death characteristic time (b) and
    proliferation ratio (c) , for our selection of 20 patients.}
  \label{fig:histos}
\end{figure}

The measured velocities are consistent with DLGGs
with values in the typical [1,4] range \cite{mandonnet08}.
The characteristics death rate times are about $\taud=1.0 \pm 0.7 ~\rm{yr}$.
The ratio between the death and proliferation rates after RT is
large, typically above 5 but with a wide distribution that can go up
to 40.


We can also check if there are some correlations between the evolution before and after RT.
From the best fit models, we 
compute the following \textit{observables}:
$V_-=\frac{dr_{mod}}{dt}(t_r^-)$) the slope just before RT, $t_{min}$
the time at which the minimum is reached, $\Delta R=r(t_r)-r(t_{min})$
the amplitude decrease at this point, and $\Delta T$ the time where
the radius is the same than at RT during the regrowth phase.

On the dataset we measure the following correlation
coefficients:
\begin{eqnarray}
   \label{eq:correl}
   <V_-,\Delta R>&=+0.42 \nonumber \\
   <{V_-,t_{min}}>&=-0.27 \\
   <{V_-,\Delta T}>&=-0.63.\nonumber
\end{eqnarray}

From the (approximate) radius evolution Eq (\ref{eq:rast}) we have
\begin{eqnarray}
   \Delta R & \simeq &v\taud~(\kr-\ln \kr-1) \nonumber\\
   t_{min}&\simeq& \taud \ln \kr \\
   \Delta T&\simeq& \taud \kr \nonumber
\end{eqnarray}
where $v$ is here the theoretical value $v=2\sqrt{D.\kappa}$.

From these expressions we expect some correlation between $v$ and $\Delta R$, an
anti-correlation between $v$ and $t_{min}$ through $\kappa$
but moderate because of the logarithm,  
and a stronger one with $\Delta T$, 
which is what is measured. We can conclude that our model reproduces
correctly the correlations observed on data before and after RT.

\section{Discussion}

DLGG are tumors that always turn into a more aggressive form after
years of indolent growth \cite{el-hateer09}. They are also
resistant to RT since they systematically recur after the end of the
treatment. Modeling their dynamics, with and without treatment, can
lead to a better understanding of their evolution and their resistance
to treatments.  
 
Here we complement a classical diffusion-proliferation model (that has
already proved its usefulness for DLGG evolution) with a model of the
RT effect, as a simple time-dependent and spatially structured death
term. The spatial dependency of the death term means that cells at the
border of the tumor are killed more than cells at the center of the
tumor. The time dependency of the death term translates into a net
proliferation term (proliferation minus death) that is time-dependent:
before RT, the net proliferation is positive, then negative during a
certain time interval after RT. When the death coefficient is smaller
than the proliferation parameter $\kappa$, the net proliferation term
is positive again and the tumor resumes its growth at the same rate
as before RT.

The first qualitative feature that our model reproduces, without RT, is the fact
that proliferating cells are situated at the border of the tumor. This
spatial effect has been observed on human tissue from DLGG: by
analyzing tissue samples from stereotaxic biopsies, it has been shown that cycling cells (or proliferating cells) are situated
at the border of the tumor \cite{gerin13}. Even if DLGG do not have a necrotic (and even hypoxic)
center as higher grade gliomas, the cell density is still higher
than normal. It is thus possible that some regions of sub-optimal oxygen
concentration develop at the center of the tumor, reducing proliferation and triggering the transformation of cells into quiescent ones. 
We will see later that this spatial organization, that our model reproduces well, is crucial in the modeling of RT action. 

Another important point is the modeling of the RT effect: although it is
certainly a complicated effect that varies among patients, we argue
that our choice of a time-dependent death rate is
biologically realistic. Tumour irradiation induces both direct and
indirect effects that could lead to tumor cell death. Direct effects
are the result of radiation-induced DNA damages in cancer cells too
important to be repaired (double strand breaks in the DNA molecules).
But RT can also induce indirect damage to DNA (via reactive oxygen species), and to the tumour microenvironment such
as vasculature. It can also trigger an immune response that can contribute to the
tumor growth control \cite{wang18, sia20}. Usually, damages cells try to
repair the damages, and can even try to go through several mitoses
before triggering their death.  
All of these process can take some time, and this is why the response
to RT can be prolonged in time. We decide to model this delayed effect
by defining a characteristic time in the death term. The choice of the
exponential function for the death term is the simplest way to
introduce a characteristic time. But it could also be
justified in an other way: the linear quadratic model  stipulates that
the survival time is an exponential function of the dose received (for a review on the linear quadratic model, see \cite{mcmahon18}).
However, the efficacy of a given dose has been measured with cell
culture in 2D. In a real tumor, it is possible that the efficacy of a
dose depends on the microenvironment. It is a well-known fact that hypoxic cells (at the
center of tumors) are more resistant to RT than normoxic cells
\cite{amberger09}. Actually, this constitutes an important limitation
to the use of RT. So,  for a given dose received, the radiation could
have a smaller effect for cells close to the poor-oxygenated center
than at the well-oxygenated border, leading to a larger survival rate.
Just after RT, the more damaged cells begin to die, quiescent cells
are now at the border and turn into proliferating ones and thus die
also, but with a death rate smaller than the first layer. And so on,
cells would die layer by layer, from the outside inwards. This process
would justify our exponentially decreasing with time death rate.
Biologically, this process is realistic: it has been actually observed
in vitro and modeled with a cellular automaton for spheroids in \cite{bruningk19, bruningk20}.

When compared to numerous high-quality clinical data (patients follow-up with tumor radius
measurements for several time points), our time dependent death rate
model, could reproduce the exponential shape visible on the
experimental data: a sharp decline in a first time, followed by a
slower decay, and an almost linear regrowth. With simple analytical considerations, we show that with a constant death term, the decrease of the
radius can only be linear (at best) and cannot lead to any
exponential-like decrease. For example, in our model of RT with
edema, the early evolution is clearly linear \cite{badoual14}.

We took care not to introduce too many parameters in out model, still allowing
flexibility to describe all the data. The tumour
evolution with RT is only described by 4 parameters, 
two being the natural evolution ones (proliferation
and diffusion) and two for the RT effect (death rate and characteristic
time).  An original aspect of this work is that we also considered the
(unknown) tumour age as a free parameter that could therefore be constrained by the
data. 

With this five-parameter model, the data of the temporal evolution of
tumors radius for 43 patients were fitted automatically and give
excellent results. We selected
20 patients with more than 10 data points (the fits for the other 23 patients are available in the Supplementary Material), and for each patient of
this series, by scanning the possible ages of the tumor, from 0 to the
patient age, we could infer the possible age range of the patient at
the onset of the tumor. We find that the age  at the onset of the 
tumor compatible with most of the patients is around 15 years-old.
This finding confirms previous research
\cite{gerin12}, where, from the data of velocity and one measure of the
tumor radius at a given time and going back in time with a model, the conclusion was
that patients were most likely to be in their late teenage years at the onset of the tumor.   

This age at the onset of the tumor depends on the
initial conditions: if the simulation starts from a small clump of
cells, the time needed to from that clump is not counted and the age
of the tumor is underestimated. Moreover, the choice of the size of
the clump would have been subjective. We chose to start the
simulations from one cell. We also assumed that the proliferation and
diffusion coefficients were constant all along the tumor evolution.
This is a strong assumption that may not be correct. It is indeed
possible that the first cell only proliferates, and form a small clump
of cells before diffusion takes place. On the other hand, since we do
not have any clue 
about what happened at the beginning of the evolution, and since even
when discovered early, DLGG seem to grow the same way as larger tumors
(associating 
proliferation and diffusion), we decided that the simplest way to
choose the initial conditions was to start with one unique cell and 
the same proliferation and diffusion coefficient. 

For the population of 20 selected patients, we also measured a RT
characteristic time $\taud$ of around 1 year and a 
ratio of the maximal death coefficient to the proliferation
coefficient $\kappa_d/\kappa$ always larger than 5. The fact that this ratio has a value larger than 1 is an important biological result: it means that just after RT, a
large quantity of cells is killed just after RT and does not have time
to go through any mitosis. In these cells, the damage due to RT may
have been so important that the cells trigger the apoptosis program
immediately, without even trying to perform a mitosis.

With the group of the 20 selected patients, we could also highlight good
correlations between the velocity before RT, $V^-$ and both the gain of
lifetime $\Delta T$ and $\Delta R$ (the maximum decrease of the radius).
A simple analytical analysis allows to
understand these correlations, that will be used, in a further work,
to perform predictions.

\appendixtitles{no} 
\appendixstart
\appendix
\section{}
\label{sec:app}
In this Appendix we compare the evolution of the tumor radius obtained
with our model, with a constant death rate model with two populations,
inspired by  Perez'  and Ribba's work \cite{ribba12, perez15}. 
In this model, RT damages a fraction of the cell population and this
damaged population evolves differently from the undamaged one: the
undamaged cell continue to proliferate and diffuse at the same rate as
before, whereas the damaged population stops proliferating and die
progressively (but still diffusing normally). A model of this type is
interesting since it is biologically realistic and it accounts for the
delay of the radius regrowth after RT. All the
parameters are constant and beside the two parameters $\kappa$ and $D$
describing the natural evolution of the tumor, two parameters are
needed for RT, the fraction of the cell population that has been
damaged, $x$, and the death rate of this population, $\kappa_d$. 

For $t<t_r$, the equation describing the evolution of the cell
population is the same as Eq (\ref{eq:freemodel}): 

$${\partial \rho \over\partial t}=D\Delta\rho+\kappa\rho(1-\rho)$$

For $t=t_r$, two population are created: the damaged one, $\rho_d(t_r)=x\rho(t_r)$ and the non damaged one : $\rho_{nd}(t_r)=(1-x)\rho_{nd}(t_r)$.
After RT ($t>t_r)$:

$${\partial \rho_{nd} \over\partial t}=D\Delta\rho_{nd}+\kappa\rho_{nd}(1-\rho)$$ for the non damaged population, and :

$${\partial \rho_{d} \over\partial t}=D\Delta\rho_{d}-\kappa_D\rho_{d}(1-\rho)$$ with $\rho=\rho_{d}+\rho_{nd}$.

This model can reproduce the regrowth delay after RT. However, as
shown in Sect.\ref{sec:carac} the
decrease of the radius is at best linear, which is as is clear from
Figure \ref{fig:bf}, 
not in agreement with the clinical data, where a steep decrease is generally first
observed followed by some milder phase. On Figure \ref{fig:comp2pop1pop},
the time of RT has been set to 0, and the RT parameters of the two
models have been chosen so that the minimum of the radius evolution is
the same. The linear decrease obtained with a two-population model
after RT can be compared to the exponential decay with our
one-population model. In the regrowth phase, the evolution of the
radius with the two-population model displays a strong curvature
reminiscent of the beginning of the visible radius evolution.     

\begin{figure}[htbp]
  \centering
  \includegraphics[width=0.7\textwidth]{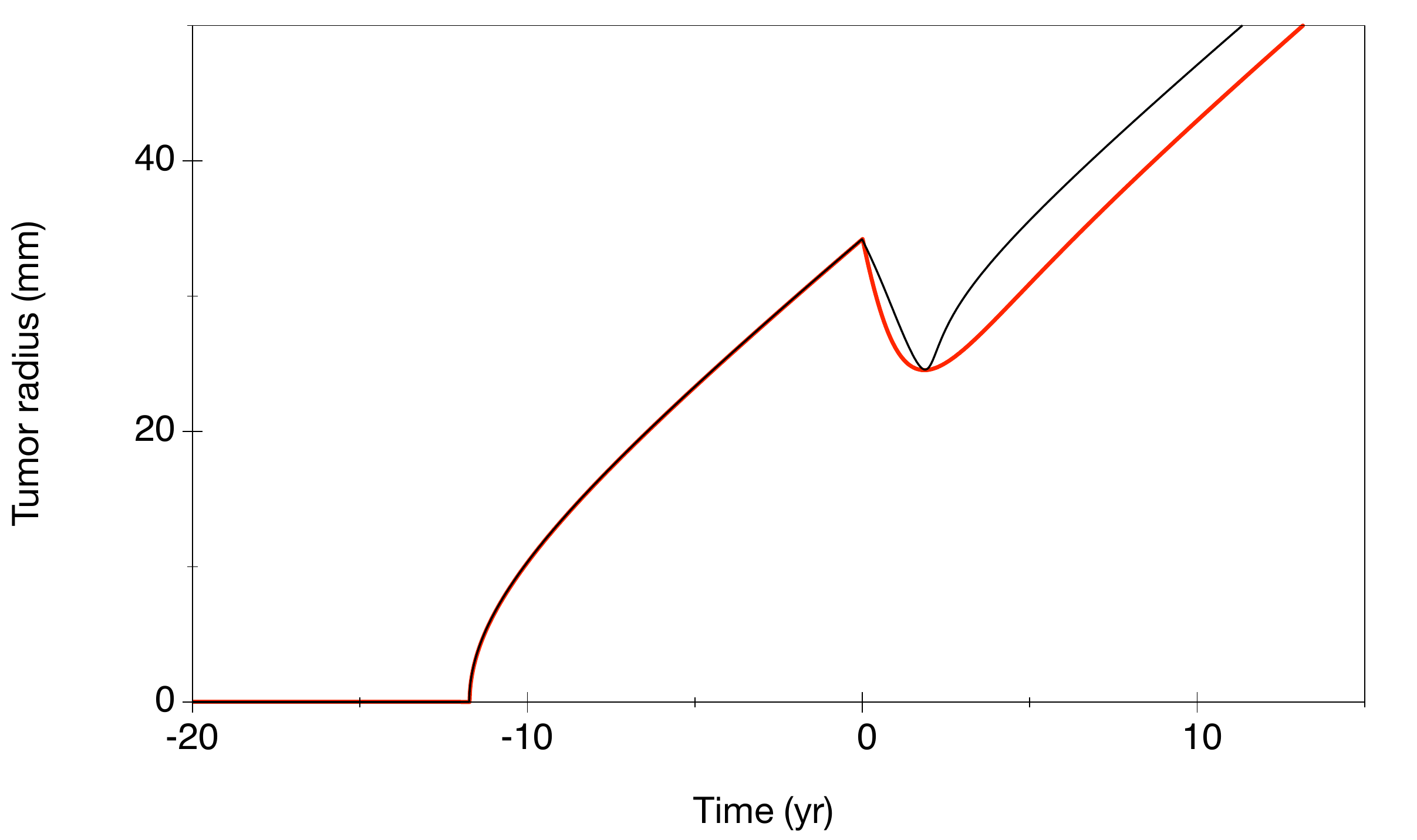}
  \caption{Temporal evolution of the tumor radius versus time, for two models, one with a constant death rate (black line) and a fraction of the cell density damaged at the time of RT and the model with a time dependent death rate model (red line). The origin of time has been set at the time of RT. Parameters for the constant death rate model : $\kappa = 1$~yr$^{-1}$, $D = 1$~mm$^2$~yr$^{-1}$, $\kappa_d=3.3$~yr$^{-1}$, $x = 0.0053$. Parameters for the time dependent death rate model : $\kappa = 1$~yr$^{-1}$, $D = 1$~mm$^2$~yr$^{-1}$, $\kappa_D=8.5$~yr$^{-1}$,$\taud=0.9$~yr.}
  \label{fig:comp2pop1pop}
\end{figure}

\end{paracol}

\reftitle{References}
\bibliography{radio}

\end{document}